\documentclass{article}
\usepackage{amssymb}
\usepackage{amsmath}

\input{tcilatex}
\begin{document}

\begin{center}

\bigskip

{\LARGE Generations of monic polynomials such that the \textit{coefficients}
of the polynomials of the next generation coincide with the \textit{zeros}
of the polynomials of the current generation, and new \textit{solvable}
many-body problems}\bigskip

$^{\ast }$\textbf{Oksana Bihun}$^{1}$ and $^{+\lozenge }$\textbf{Francesco
Calogero}$^{2}\bigskip $

$^{\ast }$Department of Mathematics, University of Colorado, Colorado
Springs, CO, USA

$^{+}$Physics Department, University of Rome \textquotedblleft La
Sapienza'', Italy

$^{\lozenge }$Istituto Nazionale di Fisica Nucleare, Sezione di Roma, Italy

$^{1}$obihun@uccs.edu

$^{2}$francesco.calogero@roma1.infn.it, francesco.calogero@uniroma1.it

\bigskip

\textit{Abstract}
\end{center}

The notion of generations of monic polynomials such that the \textit{coefficients%
} of the polynomials of the \textit{next} generation coincide with the \textit{%
zeros} of the polynomials of the \textit{current} generation is introduced, and
its relevance to the identification of endless sequences of \textit{new
solvable} many-body problems ``of goldfish type'' is demonstrated.

\bigskip


\textbf{Short title}: Generations of polynomials and solvable many-body
problems


\bigskip

\section{Introduction}

\textbf{Notation 1.1}. Unless otherwise indicated, hereafter $N$ is an 
\textit{arbitrary positive integer}, $N\geq 2$, indices such as $n,$ $m,$ $%
\ell ,$ $j,$ $...$ run over the \textit{integers} from $1$ to $N$, and
superimposed arrows denote $N$-vectors: for instance the vector $\vec{z}$
has the $N$ components $z_{n}$. Exception: hereafter $\vec{\mu}^{\left(
k\right) }$ is a $k$-vector, with each of its $k$ components $\mu _{1},$ $%
\mu _{2},...,\mu _{k}$ being \textit{integers} in the range $1\leq \mu \leq
N!$ (note that this implies that $\vec{\mu}^{\left( 1\right) }\equiv \mu
_{1} $ is a \textit{scalar}). We use instead a superimposed tilde to denote
an \textit{unordered} set of $N$ numbers: for instance the notation $\tilde{z%
}$ denotes the \textit{unordered} set of $N$ numbers $z_{n}$. Upper-case 
\textbf{boldface} letters denote $N\times N$ matrices: for instance the
matrix $\mathbf{M}$ features the $N^{2}$ elements $M_{nm}$. The numbers we
use are generally assumed to be \textit{complex} numbers; except for those
restricted to be \textit{positive integers} (see above), which generally
play the role of indices; and ``time'', see below. The \textit{imaginary unit%
} is hereafter denoted as $\mathbf{i}$, implying of course $\mathbf{i}^{2}=-1
$. For quantities depending on the \textit{real} independent variable $t$
(\textquotedblleft time\textquotedblright ) superimposed dots indicate
differentiation with respect to it: so, for instance, $\dot{z}_{n}\left(
t\right) \equiv dz_{n}\left( t\right) /dt$, $\ddot{z}_{n}\equiv
d^{2}z_{n}/dt^{2}$ (but often the $t$-dependence is not explicitly
indicated, whenever this is unlikely to cause any misunderstanding: as, for
instance, in the second formula we just wrote). The Kronecker symbol $\delta
_{nm}$ has the usual meaning: $\delta _{nm}=1$ if $n=m$, $\delta _{nm}=0$ if 
$n\neq m$; and we denote below as $\mathbf{I}$ the \textit{unit} $N\times N$
matrix the elements of which are $\delta _{nm}$. We adopt throughout the
usual convention according to which a void sum vanishes and a void product
equals unity: $\sum_{j=J}^{K}f_{j}=0,$ $\tprod\nolimits_{j=J}^{K}f_{j}=1$ if 
$J>K$. Finally we introduce the following two convenient notations: 
\begin{subequations}
\label{sigma}
\begin{equation}
\sigma _{m}\left( \vec{z}\right) =\sum_{1\leq s_{1}<s_{2}<...<s_{m}\leq
N}z_{s_{1}}z_{s_{2}}\cdots z_{s_{m}}~,  \label{sigmam}
\end{equation}%
\begin{equation}
\sigma _{n,m}\left( \vec{z}\right) =\delta _{1m}+\sum_{\substack{ 
{\footnotesize 1\leq s_{1}<s_{2}<\ldots <s_{m-1}\leq N~;}  \\ {\footnotesize %
s_{j}\neq n,~j=1,...,m-1}}}z_{s_{1}}z_{s_{2}}\cdots z_{s_{m-1}}~,
\label{sigmanm}
\end{equation}%
where of course the symbol 
\end{subequations}
\begin{equation*}
\sum_{1\leq s_{1}<s_{2}<...<s_{m}\leq N}
\end{equation*}%
denotes the sum from $1$ to $N$ over the $m$ integer indices $%
s_{1},s_{2},\ldots ,s_{m}$ with the restriction that $s_{1}<s_{2}<\ldots
<s_{m}$, while the symbol%
\begin{equation*}
\sum_{\substack{ {\footnotesize 1\leq s_{1}<s_{2}<\ldots <s_{m-1}\leq N~;} 
\\ {\footnotesize s_{j}\neq n,~j=1,...,m-1}}}
\end{equation*}%
denotes the sum from $1$ to $N$ over the $m-1$ indices $s_{1},s_{2},\ldots
,s_{m-1}$ with the restriction that $s_{1}<s_{2}<\ldots <s_{m-1}$ and
moreover the requirement that \textit{all these indices be different from }$%
n $. We note that $\sigma _{n,1}(\mathbf{z})=1$ according to the convention
(see above) that a sum over an empty set of indices equals zero. $%
\blacksquare $

\textbf{Remark 1.1}. Note that the notation $\sigma _{m}\left( \tilde{z}%
\right) $ (instead of $\sigma _{m}\left( \vec{z}\right) $) is equally meaningful, 
since this quantity, see (\ref{sigmam}), depends only on the \textit{%
symmetric sums} of the $N$ components $z_{m}$ of the $N$-vector $\vec{z},$
hence it is independent of the ordering of the $N$ elements $z_{n}$ of the 
\textit{unordered} set $\tilde{z}$. The notation $\sigma _{n,m}\left( \tilde{%
z}\right) $, instead, is not properly defined and cannot therefore be used;
except in the context of expressions which remain valid for \textit{any}
ordering of the $N$ numbers $z_{n}$, i. e., for any assignments of the $N$
different integer labels $n$ (in the range $1\leq n\leq N)$ to the $N$
elements of the \textit{unordered} set $\tilde{z}$; provided of course that the
assignment is maintained throughout that expression (in which case the
relevant expression amounts in fact to $N!$ different formulas; assuming, as
we generally do, that the $N$ numbers $z_{n}$ are \textit{all different
among themselves}). This remark is of course equally valid for any function $%
f\left( \tilde{z}\right) $. $\blacksquare $

The main protagonists of this paper are the time-dependent monic polynomials of
degree $N$ in the variable $z$: 
\begin{subequations}
\label{Pol}
\begin{equation}
p_{N}\left( z;\vec{y}\left( t\right) ;\tilde{x}\left( t\right) \right)
=z^{N}+\sum_{m=1}^{N}\left[ y_{m}\left( t\right) ~z^{N-m}\right] ~,
\label{Poly}
\end{equation}%
\begin{equation}
p_{N}\left( z;\vec{y}\left( t\right) ;\tilde{x}\left( t\right) \right)
=\prod\limits_{n=1}^{N}\left[ z-x_{n}\left( t\right) \right] ~.  \label{Polx}
\end{equation}%
Note that the notation we employ for these polynomials is somewhat
redundant, since they are equally well defined by the (time-dependent) $N$%
-vector $\vec{y}\left( t\right) $ the $N$ components of which are the $N$ 
\textit{coefficients} $y_{m}\left( t\right) $ of the polynomial (see (\ref%
{Poly})), as by the (time-dependent) unordered set $\tilde{x}\left( t\right) 
$ the $N$ elements of which are the $N$ \textit{zeros} $x_{n}\left( t\right) 
$ of the polynomial (see (\ref{Polx})). Indeed the $N$ \textit{coefficients} 
$y_{m}\left( t\right) $ can be \textit{explicitly} expressed in terms of the 
$N$ \textit{zeros }$x_{n}\left( t\right) $: 
\end{subequations}
\begin{equation}
y_{m}=\left( -1\right) ^{m}~\sigma _{m}\left( \vec{x}\right) \equiv \left(
-1\right) ^{m}~\sigma _{m}\left( \tilde{x}\right)  \label{ysigma}
\end{equation}%
(see \textbf{Notation 1.1} and \textbf{Remark 1.1}). And the $N$ \textit{%
zeros }$x_{n}\left( t\right) $ are likewise uniquely determined (up to
permutations) by the $N$ \textit{coefficients} $y_{m}\left( t\right) $, but
of course \textit{explicit} expressions to this effect are generally
available only for $N\leq 4$.

There holds moreover the following identity: 
\begin{subequations}
\label{Id12}
\begin{equation}
\left( x_{n}\right) ^{N}+\sum_{m=1}^{N}\left[ y_{m}~\left( x_{n}\right)
^{N-m}\right] =0~,  \label{Identity1}
\end{equation}%
which is an obvious consequence of (\ref{Pol}), and via (\ref{ysigma}) it
implies%
\begin{equation}
\left( x_{n}\right) ^{N}+\sum_{m=1}^{N}\left[ \left( -1\right) ^{m}~\sigma
_{m}\left( \tilde{x}\right) ~\left( x_{n}\right) ^{N-m}\right] =0~.
\label{Identity2}
\end{equation}%
Note that, while the formula (\ref{Identity1}) is an identity valid for the $%
N$ \textit{coefficients} $y_{m}$ and the $N$ \textit{zeros }$x_{n}$ of any
polynomial, see (\ref{Pol}), the identity (\ref{Identity2}) is clearly valid
for \textit{any} arbitrary assignment of the $N$ elements $x_{n}$ of the
unordered set $\tilde{x}$.

Likewise, there holds the following formula that is also clearly valid for 
\textit{any} assignment of the $N$ elements $x_{n}$ of the unordered set $%
\tilde{x}$ (see \textbf{Notation 1.1} and \textbf{Remark 1.1}): 
\end{subequations}
\begin{subequations}
\label{Id3}
\begin{equation}
-\left[ \prod\limits_{\ell =1,~\ell \neq n}^{N}\left( x_{n}-x_{\ell }\right)
^{-1}\right] ~\sum_{j=1}^{N}\left[ \left( -1\right) ^{j}~\left( x_{n}\right)
^{N-j}~\sigma _{m,j}\left( \tilde{x}\right) \right] =\delta _{nm}~;
\end{equation}%
note that this formula can also be rewritten in the following $\left(
N\times N\right) $-matrix version:%
\begin{equation}
\left[ \mathbf{R}\left( \tilde{x}\right) \right] _{nm}\equiv R_{nm}\left( 
\tilde{x}\right) =-\left[ \prod\limits_{\ell =1,~\ell \neq n}^{N}\left(
x_{n}-x_{\ell }\right) ^{-1}\right] ~\left( x_{n}\right) ^{N-m}~,  \label{R}
\end{equation}%
\begin{equation}
\left[ \mathbf{R}^{-1}\left( \tilde{x}\right) \right] _{nm}\equiv \left[
R^{-1}\left( \tilde{x}\right) \right] _{nm}=\left( -1\right) ^{n}~\sigma
_{n,m}\left( \tilde{x}\right) ~,  \label{Rinv}
\end{equation}%
implying of course (see \textbf{Notation 1.1} and \textbf{Remark 1.1}) 
\begin{equation}
\mathbf{R}\left( \tilde{x}\right) ~\mathbf{R}^{-1}\left( \tilde{x}\right) =%
\mathbf{R}^{-1}\left( \tilde{x}\right) ~\mathbf{R}\left( \tilde{x}\right) =%
\mathbf{I~.}
\end{equation}

Two more identities---the proof of which is elementary, see for instance the
Appendix  of \cite{C2015}---relate the time evolution of the $N$ \textit{%
coefficients} $y_{m}\left( t\right) $ and the $N$ \textit{zeros }$%
x_{n}\left( t\right) $ of any monic time-dependent polynomial of degree $N$
(see (\ref{Pol}), \textbf{Notation 1.1} and \textbf{Remark 1.1}): 
\end{subequations}
\begin{subequations}
\label{xydot}
\begin{equation}
\dot{x}_{n}=-\left[ \prod\limits_{\ell =1,~\ell \neq n}^{N}\left(
x_{n}-x_{\ell }\right) ^{-1}\right] ~\sum_{m=1}^{N}\left( x_{n}\right)
^{N-m}~\dot{y}_{m}~,  \label{xydota}
\end{equation}%
namely, in vector-matrix form,%
\begin{equation}
\overset{\cdot }{\tilde{x}}=\mathbf{R}\left( \tilde{x}\right) ~\overset{%
\cdot }{\vec{y}}~;  \label{xydotb}
\end{equation}%
and 
\end{subequations}
\begin{equation}
\ddot{x}_{n}=\sum_{\ell =1,~\ell \neq n}^{N}\left( \frac{2~\dot{x}_{n}~\dot{x%
}_{\ell }}{x_{n}-x_{\ell }}\right) -\left[ \prod\limits_{\ell =1,~\ell \neq
n}^{N}\left( x_{n}-x_{\ell }\right) ^{-1}\right] ~\sum_{m=1}^{N}\left(
x_{n}\right) ^{N-m}~\ddot{y}_{m}~.  \label{xydotdot}
\end{equation}

And let us report an additional identity which is an obvious consequence of
the definitions (\ref{ysigma}), (\ref{sigmam}) and (\ref{sigmanm}) (see 
\textbf{Notation 1.1} and \textbf{Remark 1.1}): 
\begin{subequations}
\label{yxdot}
\begin{equation}
\dot{y}_{m}=\left( -1\right) ^{m}~\dot{\sigma}_{m}\left( \vec{x}\right)
\equiv \left( -1\right) ^{m}~\sum_{n=1}^{N}\left[ \sigma _{n,m}\left( \tilde{%
x}\right) ~\dot{x}_{n}\right]  \label{yxdota}
\end{equation}%
or, equivalently (see (\ref{Rinv}) and (\ref{xydotb})) 
\begin{equation}
\overset{\cdot }{\vec{y}}=\mathbf{R}^{-1}\left( \tilde{x}\right) ~\overset{%
\cdot }{\tilde{x}}~.  \label{yxdotb}
\end{equation}

In the following Section 2 the notion is introduced of generations of monic
polynomials such that the \textit{coefficients} of the polynomials of the 
\textit{next} generation coincide with the \textit{zeros} of the polynomials of the 
\textit{current} generation, while Section 3 indicates how this notion is
instrumental in the identification of endless sequences of \textit{solvable} 
$N$-body problems. The paper is then concluded by a section entitled
``Outlook'', where further investigations of the generations of polynomials
introduced in Section 2, and of the \textit{solvable} $N$-body problems
identified in Section 3, are outlined.

\bigskip

\section{Generations of monic polynomials}

In this section we introduce the generations of monic polynomials such that the 
\textit{coefficients} of the polynomials of the \textit{next} generation
coincide with the \textit{zeros} of the polynomials of the \textit{current}
generation. This notion does not require that the polynomials under
consideration be time-dependent---which will instead be essential for the
developments reported in the following section; it instead requires a
certain notational generalization in the expression of the polynomials via
their \textit{coefficients} and \textit{zeros}, as shown by the following
formulas which generalize expressions (\ref{Pol}): 
\end{subequations}
\begin{subequations}
\label{Polk}
\begin{equation}
p_{N}^{\left( \vec{\mu}^{(k)}\right) }\left( z;\vec{y}^{\left( \vec{\mu}^{\left( k\right)
}\right) };\tilde{x}^{\left( \vec{\mu}^{\left( k\right) }\right) }\right)
=z^{N}+\sum_{m=1}^{N}\left[ y_{m}^{\left( \vec{\mu}^{\left( k\right)
}\right) }~z^{N-m}\right] ~,  \label{Polyk}
\end{equation}%
\begin{equation}
p_{N}^{\left( \vec{\mu}^{(k)}\right) }\left( z;\vec{y}^{\left( \vec{\mu}^{\left( k\right)
}\right) };\tilde{x}^{\left( \vec{\mu}^{\left( k\right) }\right) }\right)
=\prod\limits_{n=1}^{N}\left[ z-x_{n}^{\left( \vec{\mu}^{\left( k\right)
}\right) }\left( t\right) \right] ~.  \label{Polxk}
\end{equation}%
Here and below $k$ is a nonnegative integer taking the values $0,$ $1,$ $2,$ 
$...$, which characterizes the generations of polynomials (as explained
below), while (see \textbf{Notation 1.1}) $\vec{\mu}^{\left( k\right) }$ is
a $k$-vector the $k$ components $\mu _{1},$ $\mu _{2},...,\mu _{k}$ of which
are \textit{integers} in the range from $1$ to $N!$, the significance of
which is also explained below.

Let us start from the ``seed polynomial'' characterized by the index $k=0$,
for which we use the following notation: 
\end{subequations}
\begin{subequations}
\label{Pol0}
\begin{equation}
p_{N}^{\left( 0\right) }\left( z;\vec{y}^{\left( 0\right) };\tilde{x}%
^{\left( 0\right) }\right) =z^{N}+\sum_{m=1}^{N}\left[ y_{m}^{\left(
0\right) }~z^{N-m}\right] ~,  \label{Poly0}
\end{equation}%
\begin{equation}
p_{N}^{\left( 0\right) }\left( z;\vec{y}^{\left( 0\right) };\tilde{x}%
^{\left( 0\right) }\right) =\prod\limits_{n=1}^{N}\left[ z-x_{n}^{\left(
0\right) }\right] ~.  \label{Polx0}
\end{equation}

The polynomials of the \textit{first} generation ($k=1$) are then defined as
follows (see \textbf{Notation 1.1}): 
\end{subequations}
\begin{subequations}
\label{Pol1}
\begin{equation}
p_{N}^{\left( \mu_1\right) }\left( z;\vec{y}^{\left( \mu _{1}\right) };%
\tilde{x}^{\left( \mu _{1}\right) }\right) =z^{N}+\sum_{m=1}^{N}\left[
y_{m}^{\left( \mu _{1}\right) }~z^{N-m}\right] ~,  \label{Poly1}
\end{equation}%
\begin{equation}
p_{N}^{\left( \mu_1\right) }\left( z;\vec{y}^{\left( \mu _{1}\right) };%
\tilde{x}^{\left( \mu _{1}\right) }\right) =\prod\limits_{n=1}^{N}\left[
z-x_{n}^{\left( \mu _{1}\right) }\right] ~,  \label{Polx1}
\end{equation}%
with%
\begin{equation}
\vec{y}^{\left( \mu _{1}\right) }=\vec{x}_{\left[ \mu _{1}\right] }^{\left(
0\right) }~,  \label{yx1}
\end{equation}%
where $\vec{x}_{\left[ \mu _{1}\right] }^{\left( 0\right) }$ is the $N$%
-vector the $N$ components $x_{\left[ \mu _{1}\right] ,n}^{\left( 0\right) }$
of which coincide with the $N$ elements of the \textit{unordered} set $%
\tilde{x}^{\left( 0\right) }$---corresponding to the $N$ zeros of the 
\textit{seed} polynomial $p_{N}^{\left( 0\right) }\left( z;\vec{y}^{\left(
0\right) };\tilde{x}^{\left( 0\right) }\right) $, see (\ref{Polx0}%
)---ordered according to their permutation labeled by the specific value of
the index $\mu _{1}$ (in the range $1\leq \mu _{1}\leq N!$, since there are $%
N!$ permutations of the $N$ elements $x_{n}^{\left( 0\right) }$). Note that
this definition (\ref{Pol1}) of the first-generation (monic) polynomials
identifies in fact $N!$ different polynomials, characterized by the $N!$
values of the integer index $\mu _{1}$ in the range $1\leq \mu _{1}\leq N!$,
and that a specific \textit{unordered} set of \textit{zeros} $\tilde{x}%
^{\left( \mu _{1}\right) }$ is associated to each of them, see (\ref{Polx1}).

\textbf{Remark 2.1}. The reader who feels uneasy about the notion that a
specific permutation labeled by an index $\mu $ in the range $1\leq \mu \leq
N!$ identifies a specific order of the $N$ elements $x_{n}$ of an \textit{a
priori} \textit{unordered} set $\tilde{x}$ may reason as follows. 
A lexicographic order of the $N$  \textit{different} complex numbers $x_n$ from the \textit{%
unordered} set  $\{x_1, \ldots, x_N\}$ can be used to obtain the \textit{first} permutation 
of this unordered set. Recall that the lexicographic
rule states that, of two elements with \textit{different real} parts, the
one with \textit{algebraically smaller real} part comes \textit{first}, and
of two elements with \textit{equal real} parts, the one with \textit{%
algebraically smaller imaginary} part comes \textit{first}. Once the first permutation $\vec{x}_{[1]}$ of the unordered set $\{x_1, \ldots, x_N\}$
is established, one can use the lexicographic order of the permutations of $\vec{x}_{[1]}$ to 
index, with the index $\mu \in \{1,2, \ldots, N!\}$, the subsequent $N!$ permutations of the unordered set $\{x_1, \ldots, x_N\}$.
 Note that for simplicity we always
assume to deal with the \textit{generic} case of (monic) polynomials of
degree $N$ featuring $N$ \textit{different zeros}, and therefore as well $N$ 
\textit{different coefficients} because of the way they are defined, as now
described, see (\ref{yx1}) and below. $\blacksquare $

The next, second ($k=2$) generation of polynomials is then defined as
follows (see \textbf{Notation 1.1}): 
\end{subequations}
\begin{subequations}
\label{Pol2}
\begin{equation}
p_{N}^{\left( \vec{\mu}^{(2)}\right) }\left( z;\vec{y}^{\left( \vec{\mu}%
^{\left( 2\right) }\right) };\tilde{x}^{\left( \vec{\mu}^{\left( 2\right)
}\right) }\right) =z^{N}+\sum_{m=1}^{N}\left[ y_{m}^{\left( \vec{\mu}%
^{\left( 2\right) }\right) }~z^{N-m}\right] ~,  \label{Poly2}
\end{equation}%
\begin{equation}
p_{N}^{\left( \vec{\mu}^{(2)}\right) }\left( z;\vec{y}^{\left( \vec{\mu}%
^{\left( 2\right) }\right) };\tilde{x}^{\left( \vec{\mu}^{\left( 2\right)
}\right) }\right) =\prod\limits_{n=1}^{N}\left[ z-x_{n}^{\left( \vec{\mu}%
^{\left( 2\right) }\right) }\right] ~,  \label{Polx2}
\end{equation}%
with%
\begin{equation}
\vec{y}^{\left( \vec{\mu}^{\left( 2\right) }\right) }=\vec{x}_{\left[ \mu
_{2}\right] }^{\left( \mu _{1}\right) }~,  \label{xy2}
\end{equation}%
where $\vec{\mu}^{(2)}=(\mu_1, \mu_2)$ and $\vec{x}_{\left[ \mu _{2}\right]
}^{\left( \mu _{1}\right) }$ is the $N $-vector the $N$ components $x_{\left[
\mu _{2}\right] ,n}^{\left( \mu _{1}\right) }$ of which coincide with the $N$
elements of the \textit{unordered} set $\tilde{x}^{\left( \mu _{1}\right) }$%
---corresponding to the $N$ \textit{zeros} of the first-generation monic
polynomial $p_{N}^{\left( \mu_1\right) }\left( z;\vec{y}^{\left( \mu
_{1}\right) };\tilde{x}^{\left( \mu _{1}\right) }\right) $, see (\ref{Polx1}%
)---ordered according to their permutation labeled by the specific value of
the index $\mu _{2}$ (in the range $1\leq \mu _{2}\leq N!$, since there are $%
N!$ permutations of the $N$ elements $x_{n}^{\left( \mu _{1}\right) }$).
Note that this definition (\ref{Pol2}) of the second-generation (monic)
polynomials identifies in fact $\left( N!\right) ^{2}$ different
polynomials, characterized by the $N!$ values of the two integer indices $%
\mu _{1}$ and $\mu _{2}$ (the $2$ components of the $2$-vector $\vec{\mu}%
^{\left( 2\right) }$), each of them in the range from $1$ to $N!$, and that
a specific \textit{unordered} set of \textit{zeros} $\tilde{x}^{\left( \vec{%
\mu}^{\left( 2\right) }\right) }$ is associated to each of these
polynomials, see (\ref{Polx2}).

It is now clear how all subsequent generations of polynomials are
manufactured, all of them being generated by the initial \textit{seed}
polynomial (\ref{Pol0}). To make the matter completely clear, let us
indicate how the $k$-th generation polynomial is defined (see \textbf{%
Notation 1.1}): 
\end{subequations}
\begin{subequations}
\label{Polkk}
\begin{equation}
p_{N}^{\left( \vec{\mu}^{(k)}\right) }\left( z;\vec{y}^{\left( \vec{\mu}%
^{\left( k\right) }\right) };\tilde{x}^{\left( \vec{\mu}^{\left( k\right)
}\right) }\right) =z^{N}+\sum_{m=1}^{N}\left[ y_{m}^{\left( \vec{\mu}%
^{\left( k\right) }\right) }~z^{N-m}\right] ~,  \label{Polykk}
\end{equation}%
\begin{equation}
p_{N}^{\left( \vec{\mu}^{(k)} \right) }\left( z;\vec{y}^{\left( \vec{\mu}%
^{\left( k\right) }\right) };\tilde{x}^{\left( \vec{\mu}^{\left( k\right)
}\right) }\right) =\prod\limits_{n=1}^{N}\left[ z-x_{n}^{\left( \vec{\mu}%
^{\left( k\right) }\right) }\right] ~,  \label{Polxkk}
\end{equation}%
with%
\begin{equation}
\vec{y}^{\left( \vec{\mu}^{\left( k\right) }\right) }=\vec{x}_{\left[ \mu
_{k}\right] }^{\left( \vec{\mu}^{\left( k-1\right) }\right) }~,  \label{xyk}
\end{equation}%
where $\vec{\mu}^{(k)}=(\vec{\mu}^{(k-1)}, \mu^k)=(\mu_1, \ldots, \mu_{k-1},
\mu_k)$ and $\vec{x}_{\left[ \mu _{k}\right] }^{\left( \vec{\mu}^{\left(
k-1\right) }\right) }$ is the $N$-vector the $N$ components of which
coincide with the $N$ elements of the \textit{unordered} set $\tilde{x}%
^{\left( \vec{\mu}^{\left( k-1\right) }\right) }$---corresponding to the $N$ 
\textit{zeros} of the $\left( k-1\right) $-generation monic polynomial $%
p_{N}^{\left( \vec{\mu}^{(k-1)}\right) }\left( z;\vec{y}^{\left( \vec{\mu}%
^{\left( k-1\right) }\right) };\tilde{x}^{\left( \vec{\mu}^{\left(
k-1\right) }\right) }\right) $---ordered according to their permutation
labeled by the specific value of the index $\mu _{k}$ (in the range $1\leq
\mu _{k}\leq N!$, since there are $N!$ permutations of the $N$ elements $%
x_{n}^{\left( \vec{\mu}^{\left( k-1\right) }\right) }$). Note that this
definition (\ref{Polkk}) of the $k$-generation (monic) polynomials
identifies in fact $\left( N!\right) ^{k}$ different polynomials,
characterized by the $N!$ values of each of the $k$ integer indices $\mu
_{1},$ $\mu _{2},...,$ $\mu _{k}$ (the $k$ components of the $k$-vector $%
\vec{\mu}^{\left( k\right) }$), each of them in the range from $1$ to $N!$,
and that a specific \textit{unordered} set of \textit{zeros} $\tilde{x}%
^{\left( \vec{\mu}^{\left( k\right) }\right) }$ is associated to each of
these polynomials, see (\ref{Polxkk}) (or, equivalently, (\ref{Polxk})).

The first three generations of the monic second-degree polynomials
constructed using this procedure, starting from the generic \textit{seed}
polynomial $p_{2}^{(0)}(z)=z^{2}+bz+c$, are provided in Appendix A.

A main motivation for the introduction of these generations of polynomials
is because they turn out to be instrumental for the solution of the sequence
of \textit{new} many-body problems defined in the following section. But we
submit that---independently from this specific application---this notion
deserves further study, given its \textit{natural/unnatural} character in
the context of the theory of polynomials: indeed, a monic polynomial is
characterized by its $N$ \textit{coefficients} as well as by its $N$ \textit{%
zeros}, so a sequence of polynomials in which these two sets of numbers
exchange sequentially their roles is an \textit{intriguing} possibility;
while interchanging the \textit{coefficients} and the \textit{zeros} of
polynomials seems \textit{unorthodox} given the different nature of these
two sets of numbers, one \textit{ordered} and the other one \textit{unordered%
}.

\bigskip

\section{Solvable $N$-body problems of goldfish type}

In this section we indicate how to identify endless sequences of \textit{%
solvable} $N$-body problems which involve quite naturally the generations of
monic polynomials discussed in the preceding section. These $N$-body
problems are characterized by equations of motion of Newtonian type
(``acceleration equals force''), describing the motion in the complex $z$%
-plane of $N$ unit-mass point-particles interacting among themselves with
prescribed forces depending on their positions and their velocities. The
prototypical example is the so-called ``goldfish'' $N$-body model (for the
name see \cite{C2001a}), characterized by the equations of motion 
\end{subequations}
\begin{equation}
\ddot{x}_{n}=\sum_{\ell =1,~\ell \neq n}^{N}\left( \frac{2~\dot{x}_{n}~\dot{x%
}_{\ell }}{x_{n}-x_{\ell }}\right) ~.  \label{Gold}
\end{equation}%
A simple generalization of these equations of motion (\ref{Gold}), featuring
the arbitrary parameter $\omega $ (and reducing to (\ref{Gold}) for $\omega
=0$), reads as follows: 
\begin{subequations}
\label{SystemIsoGold}
\begin{equation}
\ddot{x}_{n}=\mathbf{i}~\omega ~\dot{x}_{n}+\sum_{\ell =1,~\ell \neq
n}^{N}\left( \frac{2~\dot{x}_{n}~\dot{x}_{\ell }}{x_{n}-x_{\ell }}\right) ~.
\label{IsoGold}
\end{equation}%
The solution of the corresponding initial-values problem is provided by the $%
N$ roots $x_{n}\equiv x_{n}\left( t\right) $ of the following, rather neat,
single \textit{algebraic} equation in the variable $z$, 
\begin{equation}
\sum_{\ell =1,~\ell \neq n}^{N}\left[ \frac{\dot{x}_{\ell }\left( 0\right) +%
\mathbf{i}~\omega ~x_{\ell }\left( 0\right) }{z-x_{\ell }\left( 0\right) }%
\right] =\frac{\mathbf{i}~\omega }{\exp \left( \mathbf{i}~\omega ~t\right) -1%
}~,  \label{SolIsoGold}
\end{equation}%
which is actually a \textit{polynomial} equation of degree $N$ in $z$, as
seen after multiplication by the product $\prod\nolimits_{n=1}^{N}\left[
z-x_{n}\left( 0\right) \right] $. Hence---whenever the parameter $\omega $
is \textit{real} and \textit{nonvanishing}, as hereafter assumed---this
model is \textit{isochronous}: \textit{all} its solutions are \textit{%
completely periodic}, with the period $T=2\pi /\left\vert \omega \right\vert 
$ (see the function in the right-hand side of (\ref{SolIsoGold})); or
possibly, due to an exchange of the particle positions after the time $T$,
with a period which is a (generally small: see \cite{GS2005}) \textit{integer%
} \textit{multiple} of $T$.

Several \textit{solvable} generalizations of the goldfish model,
characterized by Newtonian equations of motion featuring additional forces
besides those appearing in the right-hand side of (\ref{Gold}), are known:
see for instance \cite{C1978}, the two books \cite{C2001} \cite{C2008} and
references therein, the quite recent papers \cite{C2015} \cite{BC2015} \cite%
{C2015b}, and the entry ``goldfish many-body problem'' in Google or Google
Scholar.

Above and hereafter an $N$-body model is considered \textit{solvable} if the
configuration of the system at any arbitrary time $t$ can be obtained---for
given \textit{initial} data: the \textit{initial} positions and velocities
of the $N$ particles in the complex $z$-plane---by \textit{algebraic}
operations, such as finding the $N$ zeros $x_{n}\left( t\right) $ of an 
\textit{explicitly} known time-dependent polynomial of degree $N$ in $z$ (of
course such an algebraic equation can be \textit{explicitly} solved only for 
$N\leq 4$).

\textbf{Remark 3.1}. Note however that knowledge of the configuration of the 
$N$-body system at time $t$---i. e., of the (generally complex) values of
the $N$ coordinates $x_{n}\left( t\right) $ given as the \textit{unordered}
set of the $N$ zeros of a known polynomial of degree $N$ in $z$---does 
\textit{not} allow to identify the specific coordinate, say, $x_{1}\left(
t\right) $ that has evolved over time from the specific initial data $%
x_{1}\left( 0\right) ,$ $\dot{x}_{1}\left( 0\right) $; this additional
information can only be gained by following over time the evolution of the
system, either by integrating numerically the equations of motion, or by
identifying the configurations of the system (as given by the $N$ zeros of a
polynomial) at a sequence of time intervals sufficiently close to each other
so as to guarantee the identification by continuity of the trajectory of
each particle (or at least of the specific particle under consideration).
Note however that these additional operations need not be performed with
great accuracy, even when one wishes the final configuration---including the
identity of each particle---to be known with much greater accuracy.

Likewise---in the case of systems which have been identified as \textit{%
isochronous} because their solution is provided by the $N$ zeros $%
x_{n}\left( t\right) $ of a time-dependent polynomial of degree $N$ in $z$
which is itself periodic in time with period, say, $T$ (as in the above
example, see (\ref{SystemIsoGold}))---an analogous procedure must be
followed to ascertain whether the period of the time evolution of a specific
particle is $T$, or $pT$ (with $p$ a positive integer certainly not larger
than $N!$; indeed, generally much smaller, see \cite{GS2005}), due to a
periodic exchange of the correspondence between the zeros of the polynomial
and the particle identities. $\blacksquare $

The key formulas for the following developments are the identities (\ref%
{xydotdot}) and (\ref{yxdot}), relating the time evolution of the \textit{%
coefficients} $x_{n}\left( t\right) $ of the \textit{zeros} of a
time-dependent (monic) polynomial to that of the \textit{coefficients} $%
y_{m}\left( t\right) $ of the same polynomial, as well as the notion of
generations of polynomials discussed in the previous Section 2.

The starting point of our treatment is any one of the many known \textit{%
solvable} $N$-body models---see for instance \cite{C2001} \cite{C2008} and
the literature therein, and below for an example---characterized by
(Newtonian) equations of motion which we write as follows (see \textbf{%
Remark 1.1}): 
\end{subequations}
\begin{equation}
\ddot{x}_{n}=f_{n}\left( \tilde{x},~\overset{\cdot }{\tilde{x}}\right)
=f_{n}\left( \vec{x},~\overset{\cdot }{\vec{x}}\right) ~.  \label{SolvNewt}
\end{equation}%
We then consider the generations of (time-dependent) polynomials of degree $N
$ in $z$ originating from the (time-dependent) \textit{seed} polynomial 
\begin{subequations}
\label{Pol0t}
\begin{equation}
p_{N}^{\left( 0\right) }\left( z;\vec{y}^{\left( 0\right) }\left( t\right) ;%
\tilde{x}^{\left( 0\right) }\left( t\right) \right) =z^{N}+\sum_{m=1}^{N}%
\left[ y_{m}^{\left( 0\right) }\left( t\right) ~z^{N-m}\right] ~,
\end{equation}%
\begin{equation}
p_{N}^{\left( 0\right) }\left( z;\vec{y}^{\left( 0\right) }\left( t\right) ;%
\tilde{x}^{\left( 0\right) }\left( t\right) \right) =\prod\limits_{n=1}^{N}%
\left[ z-x_{n}^{\left( 0\right) }\left( t\right) \right] ~,
\end{equation}%
the (time-dependent) \textit{zeros} $x_{n}^{\left( 0\right) }\left( t\right) 
$ of which are the solution of the \textit{solvable} model (\ref{SolvNewt}):%
\begin{equation}
x_{n}^{\left( 0\right) }\left( t\right) =x_{n}\left( t\right) ~.
\label{x0xt}
\end{equation}%
It is then plain via (\ref{xydotdot}) that the zeros $x_{n}^{\left( \mu
_{1}\right) }\left( t\right) $ of the first generation polynomials (see (\ref%
{yx1}) and (\ref{x0xt})) 
\end{subequations}
\begin{subequations}
\label{Pol1t}
\begin{equation}
p_{N}^{\left( \mu _{1}\right) }\left( z;\vec{x}_{\left[ \mu _{1}\right]
}\left( t\right) ;\tilde{x}^{\left( \mu _{1}\right) }\left( t\right) \right)
=z^{N}+\sum_{m=1}^{N}\left[ x_{\left[ \mu _{1}\right] ,m}\left( t\right)
~z^{N-m}\right] ~,
\end{equation}%
\begin{equation}
p_{N}^{\left( \mu _{1}\right) }\left( z;\vec{x}_{\left[ \mu _{1}\right]
}\left( t\right) ;\tilde{x}^{\left( \mu _{1}\right) }\left( t\right) \right)
=\prod\limits_{n=1}^{N}\left[ z-x_{n}^{\left( \mu _{1}\right) }\left(
t\right) \right] ~,
\end{equation}%
provide the solutions of the $N$-body problems 
\end{subequations}
\begin{subequations}
\label{NewModels1}
\begin{eqnarray}
&&\ddot{x}_{n}^{\left( \mu _{1}\right) }=\sum_{\ell =1,~\ell \neq
n}^{N}\left( \frac{2~\dot{x}_{n}^{\left( \mu _{1}\right) }~\dot{x}_{\ell
}^{\left( \mu _{1}\right) }}{x_{n}^{\left( \mu _{1}\right) }-x_{\ell
}^{\left( \mu _{1}\right) }}\right)   \notag \\
&&-\left[ \prod\limits_{\ell =1,~\ell \neq n}^{N}\left( x_{n}^{\left( \mu
_{1}\right) }-x_{\ell }^{\left( \mu _{1}\right) }\right) ^{-1}\right]
~\sum_{m=1}^{N}\left( x_{n}^{\left( \mu _{1}\right) }\right)
^{N-m}~f_{m}\left( \vec{y}^{\left( \mu _{1}\right) },~\dot{\vec{y%
}}^{\left( \mu _{1}\right) }\right)   \notag \\
&&
\end{eqnarray}%
with, in the right hand side, the components of the $N$-vectors $\vec{y}%
^{\left( \mu _{1}\right) }\equiv \vec{y}^{\left( \mu _{1}\right) }\left(
t\right) $ and $\dot{\vec{y}}^{\left( \mu _{1}\right) }\equiv d%
\vec{y}^{\left( \mu _{1}\right) }\left( t\right) /dt$ replaced of course by
their expressions in terms of the components $x_{\left[ \mu _{1}\right]
,n}\left( t\right) $ respectively $\dot{x}_{\left[ \mu _{1}\right] ,n}\left(
t\right) $ of the $N$-vectors $\vec{x}_{\left[ \mu _{1}\right] }\left(
t\right) $ respectively $d\vec{x}_{\left[ \mu _{1}\right] }\left( t\right)
/dt$ as follows:%
\begin{equation}
y_{m}^{\left( \mu _{1}\right) }=\left( -1\right) ^{m}~\sigma _{m}\left( \vec{%
x}_{\left[ \mu _{1}\right] }\right) ~,
\end{equation}%
\begin{equation}
\dot{y}_{m}^{\left( \mu _{1}\right) }=\left( -1\right) ^{m}~\sum_{n=1}^{N}%
\left[ \sigma _{n,m}\left( \vec{x}_{\left[ \mu _{1}\right] }\right) ~\dot{x}%
_{n}^{\left( \mu _{1}\right) }\right] 
\end{equation}%
(see (\ref{ysigma}) and (\ref{yxdot})). Note that in this manner we have
identified $N!$ new $N$-body problems, labeled by the index $\mu _{1}$
taking integer values in the range from $1$ to $N!$ (the significance of
which is explained in Section 2), and characterized by the Newtonian
equations of motion of goldfish type (\ref{NewModels1}). These \textit{new} $%
N$-body problems are of course \textit{solvable}, since their solutions are
provided by the \textit{zeros} of the polynomials (\ref{Pol1t}) which are
known because their \textit{coefficients} are (a given permutation,
characterized by the index $\mu _{1}$, of) the solutions of the problem (\ref%
{SolvNewt}), which is assumed to begin with to be itself \textit{solvable}.

And it is now plain how this technique can be iterated over and over again
in order to identify \textit{new} \textit{solvable} $N$-body problems. Let
us just exhibit---relying on the notation of Section 2---the $\left(
N!\right) ^{2}$ \textit{solvable} $N$-body problems yielded by the first
iteration of this procedure. The corresponding Newtonian equations of motion
of goldfish type read as follows: 
\end{subequations}
\begin{subequations}
\label{NewModels2}
\begin{eqnarray}
&&\ddot{x}_{n}^{\left( \vec{\mu}^{\left( 2\right) }\right) }=\sum_{\ell
=1,~\ell \neq n}^{N}\left( \frac{2~\dot{x}_{n}^{\left( \vec{\mu}^{\left(
2\right) }\right) }~\dot{x}_{\ell }^{\left( \vec{\mu}^{\left( 2\right)
}\right) }}{x_{n}^{\left( \vec{\mu}^{\left( 2\right) }\right) }-x_{\ell
}^{\left( \vec{\mu}^{\left( 2\right) }\right) }}\right)   \notag \\
&&-\left[ \prod\limits_{\ell =1,~\ell \neq n}^{N}\left( x_{n}^{\left( \vec{%
\mu}^{\left( 2\right) }\right) }-x_{\ell }^{\left( \vec{\mu}^{\left(
2\right) }\right) }\right) ^{-1}\right] ~\sum_{m=1}^{N}\left( x_{n}^{\left( 
\vec{\mu}^{\left( 2\right) }\right) }\right) ^{N-m}~\ddot{y}_{m}^{\left( 
\vec{\mu}^{\left( 2\right) }\right) }~,  \notag \\
&&  \label{NewModels2a}
\end{eqnarray}%
where $\vec{\mu}^{\left( 2\right) }$ is the $2$-vector with two components $%
\mu _{1}$ and $\mu _{2}$ (each of them taking integer values from $1$ to $N!$%
). For each of the $\left( N!\right) ^{2}$ assignment of this $2$-vector $%
\vec{\mu}^{\left( 2\right) }$ (the significance of which has been explained
in Section 2), this set of $N$ Newtonian equations of motion of goldfish
type determine the time-evolution of the coordinates $x_{n}^{\left( \vec{\mu}%
^{\left( 2\right) }\right) }\equiv x_{n}^{\left( \vec{\mu}^{\left( 2\right)
}\right) }\left( t\right) $, with the quantity $\ddot{y}_{m}^{\left( \vec{\mu%
}^{\left( 2\right) }\right) }$ appearing in the right-hand side of (\ref%
{NewModels2a}) being replaced as follows: 
\begin{eqnarray}
&&\ddot{y}_{m}^{\left( \vec{\mu}^{\left( 2\right) }\right) }=\sum_{\ell
=1,~\ell \neq n}^{N}\left( \frac{2~\dot{y}_{m}^{\left( \vec{\mu}^{\left(
2\right) }\right) }~\dot{y}_{\ell }^{\left( \vec{\mu}^{\left( 2\right)
}\right) }}{y_{m}^{\left( \vec{\mu}^{\left( 2\right) }\right) }-y_{\ell
}^{\left( \vec{\mu}^{\left( 2\right) }\right) }}\right)   \notag \\
&&-\left[ \prod\limits_{\ell =1,~\ell \neq m}^{N}\left( y_{m}^{\left( \vec{%
\mu}^{\left( 2\right) }\right) }-y_{\ell }^{\left( \vec{\mu}^{\left(
2\right) }\right) }\right) ^{-1}\right] ~\sum_{n=1}^{N}\left( y_{m}^{\left( 
\vec{\mu}^{\left( 2\right) }\right) }\right) ^{N-n}~f_{n}\left( \vec{y}%
^{\left( \vec{\mu}^{\left( 2\right) }\right) },~\dot{\vec{y}}%
^{\left( \vec{\mu}^{\left( 2\right) }\right) }\right) ~,  \notag \\
&&
\end{eqnarray}%
where moreover, in the right-hand side of this expression, the quantities $%
y_{m}^{\left( \vec{\mu}^{\left( 2\right) }\right) }\equiv y_{m}^{\left( \vec{%
\mu}^{\left( 2\right) }\right) }\left( t\right) $ and $\dot{y}_{m}^{\left( 
\vec{\mu}^{\left( 2\right) }\right) }=\dot{y}_{m}^{\left( \vec{\mu}^{\left(
2\right) }\right) }\left( t\right) $ should be replaced by their expressions
in terms of the coordinates $x_{n}^{\left( \vec{\mu}^{\left( 2\right)
}\right) }\equiv x_{n}^{\left( \vec{\mu}^{\left( 2\right) }\right) }\left(
t\right) $ and their time-derivatives $\dot{x}_{n}^{\left( \vec{\mu}^{\left(
2\right) }\right) }\equiv \dot{x}_{n}^{\left( \vec{\mu}^{\left( 2\right)
}\right) }\left( t\right) ,$ as follows (see again (\ref{ysigma}) and (\ref%
{yxdot})):%
\begin{equation}
y_{m}^{\left( \vec{\mu}^{\left( 2\right) }\right) }=\left( -1\right)
^{m}~\sigma _{m}\left( \vec{x}^{\left( \vec{\mu}^{\left( 2\right) }\right)
}\right) ~,  \label{Replyx2}
\end{equation}%
\begin{equation}
\dot{y}_{m}^{\left( \vec{\mu}^{\left( 2\right) }\right) }=\left( -1\right)
^{m}~\sum_{n=1}^{N}\left[ \sigma _{n,m}\left( \vec{x}^{\left( \vec{\mu}%
^{\left( 2\right) }\right) }\right) ~\dot{x}_{n}^{^{\left( \vec{\mu}^{\left(
2\right) }\right) }}\right] ~.  \label{Replyx2dot}
\end{equation}

As a particularly simple example, let us display the \textit{solvable} $N$%
-body problems written above, corresponding to the assignment (with $a$ an
arbitrary constant) 
\end{subequations}
\begin{subequations}
\label{Example0}
\begin{equation}
f_{n}\left( \vec{x},~\overset{\cdot }{\vec{x}}\right) =\left( \mathbf{i}%
-a\right) ~\dot{x}_{n}-\mathbf{i}~a~x_{n}~,
\end{equation}%
implying (see (\ref{SolvNewt}))%
\begin{equation}
x_{n}\left( t\right) =\frac{\left[ a~x_{n}\left( 0\right) +\dot{x}_{n}\left(
0\right) \right] ~\exp \left( \mathbf{i~}t\right) +\left[ \mathbf{i}%
~x_{n}\left( 0\right) -\dot{x}_{n}\left( 0\right) \right] ~\exp \left(
-a~t\right) }{\mathbf{i}+a}~.
\end{equation}

Then the first generation of \textit{solvable} $N$-body problems is
characterized by the following Newtonian equations of motion of goldfish
type (see (\ref{NewModels1})) 
\end{subequations}
\begin{subequations}
\label{Example1}
\begin{eqnarray}
&&\ddot{x}_{n}^{\left( \mu _{1}\right) }=\sum_{\ell =1,~\ell \neq
n}^{N}\left( \frac{2~\dot{x}_{n}^{\left( \mu _{1}\right) }~\dot{x}_{\ell
}^{\left( \mu _{1}\right) }}{x_{n}^{\left( \mu _{1}\right) }-x_{\ell
}^{\left( \mu _{1}\right) }}\right)  \notag \\
&&-\left[ \prod\limits_{\ell =1,~\ell \neq n}^{N}\left( x_{n}^{\left( \mu
_{1}\right) }-x_{\ell }^{\left( \mu _{1}\right) }\right) ^{-1}\right]
~\sum_{m=1}^{N}\left( x_{n}^{\left( \mu _{1}\right) }\right) ^{N-m}~\left[
\left( \mathbf{i}-a\right) ~\dot{y}_{m}^{\left( \mu _{1}\right) }-\mathbf{i}%
~a~y_{m}^{\left( \mu _{1}\right) }\right]  \notag \\
&&
\end{eqnarray}%
which, via (\ref{xydota}) and (\ref{Identity1}), may in this case be
simplified to read%
\begin{eqnarray}
&&\ddot{x}_{n}^{\left( \mu _{1}\right) }=\sum_{\ell =1,~\ell \neq
n}^{N}\left( \frac{2~\dot{x}_{n}^{\left( \mu _{1}\right) }~\dot{x}_{\ell
}^{\left( \mu _{1}\right) }}{x_{n}^{\left( \mu _{1}\right) }-x_{\ell
}^{\left( \mu _{1}\right) }}\right) +\left( \mathbf{i}-a\right) ~\dot{x}%
_{n}^{\left( \mu _{1}\right) }  \notag \\
&&-\mathbf{i}~a~\left[ \prod\limits_{\ell =1,~\ell \neq n}^{N}\left(
x_{n}^{\left( \mu _{1}\right) }-x_{\ell }^{\left( \mu _{1}\right) }\right)
^{-1}\right] ~\left( x_{n}^{\left( \mu _{1}\right) }\right) ^{N}~.
\end{eqnarray}

While the equations of motion of the Newtonian equations of motion of
goldfish type characterizing the second generation of \textit{solvable} $N$%
-body problems read (see (\ref{NewModels2})) 
\end{subequations}
\begin{subequations}
\label{Example2}
\begin{eqnarray}
&&\ddot{x}_{n}^{\left( \vec{\mu}^{\left( 2\right) }\right) }=\sum_{\ell
=1,~\ell \neq n}^{N}\left( \frac{2~\dot{x}_{n}^{\left( \vec{\mu}^{\left(
2\right) }\right) }~\dot{x}_{\ell }^{\left( \vec{\mu}^{\left( 2\right)
}\right) }}{x_{n}^{\left( \vec{\mu}^{\left( 2\right) }\right) }-x_{\ell
}^{\left( \vec{\mu}^{\left( 2\right) }\right) }}\right)   \notag \\
&&-\left[ \prod\limits_{\ell =1,~\ell \neq n}^{N}\left( x_{n}^{\left( \vec{%
\mu}^{\left( 2\right) }\right) }-x_{\ell }^{\left( \vec{\mu}^{\left(
2\right) }\right) }\right) ^{-1}\right] ~\sum_{m=1}^{N}\left( x_{n}^{\left( 
\vec{\mu}^{\left( 2\right) }\right) }\right) ^{N-m}~\ddot{y}_{m}^{\left( 
\vec{\mu}^{\left( 2\right) }\right) }~,  \notag \\
&&  \label{Example2a}
\end{eqnarray}%
with the quantity $\ddot{y}_{m}^{\left( \vec{\mu}^{\left( 2\right) }\right) }
$ appearing in the right-hand side of (\ref{Example2a}) being replaced as
follows: 
\begin{eqnarray}
&&\ddot{y}_{m}^{\left( \vec{\mu}^{\left( 2\right) }\right) }=\sum_{\ell
=1,~\ell \neq n}^{N}\left( \frac{2~\dot{y}_{m}^{\left( \vec{\mu}^{\left(
2\right) }\right) }~\dot{y}_{\ell }^{\left( \vec{\mu}^{\left( 2\right)
}\right) }}{y_{m}^{\left( \vec{\mu}^{\left( 2\right) }\right) }-y_{\ell
}^{\left( \vec{\mu}^{\left( 2\right) }\right) }}\right) +\left( \mathbf{i}%
-a\right) ~\dot{y}_{m}^{\left( \vec{\mu}^{\left( 2\right) }\right) }  \notag
\\
&&-\mathbf{i}~a~\left[ \prod\limits_{\ell =1,~\ell \neq m}^{N}\left(
y_{m}^{\left( \vec{\mu}^{\left( 2\right) }\right) }-y_{\ell }^{\left( \vec{%
\mu}^{\left( 2\right) }\right) }\right) ^{-1}\right] ~\left( y_{m}^{\left( 
\vec{\mu}^{\left( 2\right) }\right) }\right) ^{N}~~,  \notag \\
&&
\end{eqnarray}%
of course with the additional replacements (\ref{Replyx2}) and (\ref%
{Replyx2dot}).

\textbf{Remark 3.2}. Clearly the original $N$-body system of this example,
see (\ref{SolvNewt}) with (\ref{Example0}), has the property to be \textit{%
isochronous} (with period $2\pi $) if $a=0$, and to be \textit{%
asymptotically isochronous} if the parameter $a$ is an arbitrary \textit{%
positive} number (for the notion of \textit{asymptotic isochrony} see \cite%
{CG2008} and Chapter 6 of \cite{C2008}). And it is plain that these features
are then preserved by \textit{all} the \textit{solvable} $N$-body problems
generated from this original model (of which the first two instances are
exhibited above); and also that this property---the inheritance of \textit{%
isochrony} or \textit{asymptotic isochrony}, as the case may be, by \textit{%
all} $N$-body problems generated by an original model possessing these
features---is a general characteristic of the class of \textit{solvable}
systems generated by an iteration of the procedure based on the repeated use
of the identity (\ref{xydotdot}). $\blacksquare $

Several new solvable $N$-body problems generated by this technique---but
only restricted to the first application of the identity (\ref{xydotdot}),
without any further iteration---have been already investigated: see \cite%
{C2015} \cite{BC2015} \cite{C2015b}. In the present paper we limit our
presentation to the general introduction of this technique---which has
clearly the potential to yield endless sequences of solvable systems---and
to the single example of its application including just one iteration, as
described above. Specific discussion of other such models---as well as more
complete analyses of the behavior of the solutions of the new models
discussed in the recent papers we just quoted---might be undertaken by
ourselves or others to the extent that these findings evoke sufficient
interest.

Let us end this section with the following

\textbf{Remark 3.3}. It is clear from our treatment that the endless
sequences of \textit{solvable} $N$-body problems of goldfish type associated
via the technique described above to any \textit{solvable} \textit{seed}
problem are in fact yielded by an appropriate sequence of \textit{changes of
dependent variables}. It might therefore be concluded that all these models
are, as it were, \textit{trivially equivalent} to the original \textit{seed}
model. But such an opinion would clash with the fact that most---perhaps
all---the \textit{solvable} $N$-body models which have been identified and
investigated worldwide in the last few decades---their discovery and
analysis constituting a substantial development of mathematics and
mathematical physics over the last few decades---are as well reducible to
altogether \textit{trivial time evolutions} by \textit{appropriate changes
of dependent variables}. The rub is the identification---and the
investigation---of the \textit{appropriate changes of dependent variables};
namely, in the present context, further study of the notion of generations
of polynomials as described in Section 2. $\blacksquare $

\bigskip

\section{Outlook}

The findings reported in this paper suggest further developments, which
ourselves or others will hopefully pursue and report in future publications.

Investigation of the properties of the \textit{zeros} of the polynomials
belonging to subsequent generations as defined in Section 2---and of the
associated \textit{Riemann surfaces}---is a topic that naturally belongs to
algebraic geometry, and which does not seem to have been investigated so far
(to the best of our knowledge), while, in our opinion, it deserves further
study; as well as other properties of these polynomials, for instance when
they originate from ``named'' polynomial \textit{seeds}.

We originally believed that \textit{Diophantine} properties of the \textit{%
zeros} of the polynomials belonging to the generations of polynomials
yielded, say, by one of the classical polynomials playing the role of 
\textit{seed}, could be obtained by an appropriate investigation of
appropriate dynamical systems---of the type discussed in Section 3--- in the
immediate vicinity of their equilibria; indeed this is one of the
techniques---see for instance \cite{ABCOP1979} or \cite{C2001} \cite{C2008}
and references therein---that allows to show, for instance, that the $\left(
N\times N\right) $-matrix $\mathbf{M}\left( \tilde{x}\right) $ defined
componentwise as follows in terms of the zeros $x_{n}$ of the Hermite
polynomial $H_{N}\left( z\right) $ (see for instance \cite{E1953}) 
\end{subequations}
\begin{subequations}
\label{MHermite}
\begin{equation}
M_{nm}\left( \tilde{x}\right) =-\left( x_{n}-x_{m}\right) ^{-2}~,~~~n\neq m~,
\end{equation}%
\begin{equation}
M_{nn}\left( \tilde{x}\right) =\sum_{\ell =1,~\ell \neq n}^{N}\left(
x_{n}-x_{\ell }\right) ^{-2}=-\sum_{\ell =1,~\ell \neq n}^{N}M_{n\ell
}\left( \tilde{x}\right) ~,
\end{equation}%
features the $N$ eigenvalues $0,1,...,N-1$. (Note that these formulas define
in fact $N!$ different matrices due to the \textit{unordered} character of
the set $\tilde{x},$ but these matrices are related to each other by a
permutation of their $N$ lines and a corresponding permutation of their $N$
columns, so that the fact that they all feature the \textit{same} spectrum
is automatically guaranteed). And indeed we were able to identify in this
manner $\left( N\times N\right) $-matrices $\mathbf{M}^{\left( \mu
_{1}\right) }\left( \tilde{x},\tilde{x}^{\left( \mu _{1}\right) }\right) $
which feature the \textit{same} eigenvalues $0,1,...,N-1$, being defined in
terms of the $N$ \textit{zeros} $x_{n}$ of the Hermite polynomial $%
H_{N}\left( z\right) $ and of the $N$ \textit{zeros} $x_{n}^{\left( \mu
_{1}\right) }$ of each of the first generation of polynomials yielded, via
the technique of Section 2, by the Hermite polynomial used as \textit{seed}%
---i. e., the \textit{zeros} $x_{n}^{\left( \mu _{1}\right) }$ of the
polynomials the \textit{coefficients} of which are the \textit{zeros} of the
Hermite polynomial $H_{N}\left( z\right) $. But we also discovered that
these matrices were of the following type: 
\end{subequations}
\begin{equation}
\mathbf{M}^{\left( \mu _{1}\right) }\left( \tilde{x},\tilde{x}^{\left( \mu
_{1}\right) }\right) =\mathbf{R}\left( \tilde{x}^{\left( \mu _{1}\right)
}\right) ~\mathbf{M}\left( \tilde{x}\right) ~\left[ \mathbf{R}\left( \tilde{x%
}^{\left( \mu _{1}\right) }\right) \right] ^{-1}~,  \label{M1}
\end{equation}%
so that their property to have the same spectrum as the matrix $\mathbf{M}%
\left( \tilde{x}\right) $ implies no special property of the $N$ zeros of
the set $\tilde{x}^{\left( \mu _{1}\right) }$; indeed, this conclusion is
implied by this formula for any arbitrary definition of the matrix $\mathbf{R%
}\left( \tilde{x}^{\left( \mu _{1}\right) }\right) ,$ under the sole
condition that this matrix be invertible. And we also convinced ourselves
(see Appendix B, where the derivation of the above formula is reported)
that---at least via the technique we use there---the same phenomenon would
also happen for matrices constructed with the \textit{zeros} of subsequent
generations of polynomials or from the \textit{zeros} of polynomials
generated by different \textit{seeds}. So the identification of \textit{Diophantine}, or other, remarkable properties of the 
\textit{zeros} of the polynomials belonging to generations of polynomials of
the type yielded by the approach described in Section 2 remains an
open problem.

\bigskip

\section{Appendix A}

In this Appendix A we report the first three generations of the monic
second-degree polynomials, constructed using the procedure described in
Section~2, starting from the generic \textit{seed} polynomial 
\begin{equation}
p_{2}^{(0)}(z)=z^{2}+bz+c,  \label{N2seed}
\end{equation}%
where $b$ and $c$ are two \textit{arbitrary complex} numbers.

Before we proceed, let us note that every \textit{complex} number $\zeta
=\rho ~\exp \left( \mathbf{i~}\phi \right) $---where of course $\rho $ is a 
\textit{nonnegative real} number and $\phi $ is a \textit{real} number in
the range $0\leq \phi <2\pi $---has two square roots $\pm r$, where $r=\sqrt{%
\rho }~\exp \left( \mathbf{i~}\phi /2\right) $ and, for definiteness, the
square root $\sqrt{\rho }$ is \textit{nonnegative}. Using this fact, we
introduce the (generally complex) numbers $%
r_{0},r_{11},r_{12},r_{21},r_{22},r_{23},r_{24}$ as follows: 
\begin{subequations}
\label{rr}
\begin{eqnarray}
&&\sqrt{b^{2}-4c}=\pm r_{0},  \label{r0} \\
&&\sqrt{8b+2b^{2}-4c+8r_{0}-2br_{0}}=\pm r_{11},  \label{r11} \\
&&\sqrt{8b+2b^{2}-4c-8r_{0}+2br_{0}}=\pm r_{12},  \label{r12} \\
&&\sqrt{-8b+4b^{2}-8c+24r_{0}-4br_{0}+16r_{11}+2br_{11}-2r_{0}r_{11}}=\pm
r_{21},  \label{r21} \\
&&\sqrt{-8b+4b^{2}-8c+24r_{0}-4br_{0}-16r_{11}-2br_{11}+2r_{0}r_{11}}=\pm
r_{22},  \label{22} \\
&&\sqrt{-8b+4b^{2}-8c-24r_{0}+4br_{0}+16r_{12}+2br_{12}+2r_{0}r_{12}}=\pm
r_{23},  \label{r23} \\
&&\sqrt{-8b+4b^{2}-8c-24r_{0}+4br_{0}-16r_{12}-2br_{12}-2r_{0}r_{12}}=\pm
r_{24}.  \label{r24}
\end{eqnarray}%
\end{subequations}
Using this notation, we display below the polynomials in the first three
generations, yielded by the \textit{seed} polynomial (\ref{N2seed}).

\textit{First generation}. The polynomials in the first generation are given
by 
\begin{subequations}
\begin{eqnarray}
p_{2}^{(1)}(z) &=&z^{2}-\frac{b}{2}(z+1)+\frac{r_{0}}{2}(z-1), \\
p_{2}^{(2)}(z) &=&z^{2}-\frac{b}{2}(z+1)-\frac{r_{0}}{2}(z-1).
\end{eqnarray}

\textit{Second generation}. The polynomials in the second generation are
given by 
\end{subequations}
\begin{subequations}
\begin{eqnarray}
p_{2}^{(1,1)}(z) &=&z^{2}+\frac{b-r_{0}}{4}(z+1)+\frac{r_{11}}{4}(z-1), \\
p_{2}^{(1,2)}(z) &=&z^{2}+\frac{b-r_{0}}{4}(z+1)-\frac{r_{11}}{4}(z-1), \\
p_{2}^{(2,1)}(z) &=&z^{2}+\frac{b+r_{0}}{4}(z+1)+\frac{r_{12}}{4}(z-1), \\
p_{2}^{(2,2)}(z) &=&z^{2}+\frac{b+r_{0}}{4}(z+1)-\frac{r_{12}}{4}(z-1).
\end{eqnarray}

\textit{Third generation}. The polynomials in the third generation are given
by 
\end{subequations}
\begin{subequations}
\begin{eqnarray}
p_{2}^{(1,1,1)}(z) &=&z^{2}+\frac{1}{8}(-b+r_{0}-r_{11})(z+1)+\frac{r_{21}}{8%
}(z-1), \\
p_{2}^{(1,1,2)}(z) &=&z^{2}+\frac{1}{8}(-b+r_{0}-r_{11})(z+1)-\frac{r_{21}}{8%
}(z-1), \\
p_{2}^{(1,2,1)}(z) &=&z^{2}+\frac{1}{8}(-b+r_{0}+r_{11})(z+1)+\frac{r_{22}}{8%
}(z-1), \\
p_{2}^{(1,2,2)}(z) &=&z^{2}+\frac{1}{8}(-b+r_{0}+r_{11})(z+1)-\frac{r_{22}}{8%
}(z-1), \\
p_{2}^{(2,1,1)}(z) &=&z^{2}+\frac{1}{8}(-b-r_{0}-r_{12})(z+1)+\frac{r_{23}}{8%
}(z-1), \\
p_{2}^{(2,1,2)}(z) &=&z^{2}+\frac{1}{8}(-b-r_{0}-r_{12})(z+1)-\frac{r_{23}}{8%
}(z-1), \\
p_{2}^{(2,2,1)}(z) &=&z^{2}+\frac{1}{8}(-b-r_{0}+r_{12})(z+1)+\frac{r_{24}}{8%
}(z-1), \\
p_{2}^{(2,2,2)}(z) &=&z^{2}+\frac{1}{8}(-b-r_{0}+r_{12})(z+1)-\frac{r_{24}}{8%
}(z-1).
\end{eqnarray}

\bigskip

\section{Appendix B}

In this Appendix B we show that
the following, rather standard, approach, does not allow to
obtain results for the zeros of the polynomials of the generations yielded
by any given polynomial playing the role of the \textit{seed}.

Take as starting point a \textit{solvable} dynamical system 
\end{subequations}
\begin{subequations}
\begin{equation}
\dot{\gamma}_{m}\left( t\right) =f_{m}\left( \vec{\gamma}\left( t\right)
\right) ~,~~~\overset{\cdot }{\vec{\gamma}}=\vec{f}\left( \vec{\gamma}%
\right) ~;  \label{EqMotSolvSyst}
\end{equation}%
and assume that this system has the equilibrium configuration%
\begin{equation}
\vec{\gamma}\left( t\right) =\vec{\gamma}\left( 0\right) =\vec{x}~,
\end{equation}%
where of course%
\begin{equation}
f_{m}\left( \vec{x}\right) =0~,~~~\vec{f}\left( \vec{x}\right) =\vec{0}~.
\end{equation}

Because this system is \textit{solvable}, one can control the behavior of
its solutions for all time from all initial data. Let us then look at its
behavior infinitesimally close to its equilibrium configuration, by setting 
\end{subequations}
\begin{subequations}
\begin{equation}
\vec{\gamma}\left( t\right) =\vec{x}+\varepsilon ~\vec{w}\left( t\right) +%
\vec{o}\left( \varepsilon \right) ~,  \label{gammaw}
\end{equation}%
implying 
\begin{equation}
\overset{\cdot }{\vec{w}}\left( t\right) =\mathbf{F}\left( \vec{x}\right) 
\mathbf{~}\vec{w}\left( t\right) ~,
\end{equation}%
where the $N\times N$ time-independent matrix $\mathbf{F}\left( \vec{x}%
\right) $ is defined componentwise as follows:%
\begin{equation}
F_{nm}\left( \vec{x}\right) =\left. \frac{\partial ~f_{n}\left( \vec{\gamma}%
\right) }{\partial ~\gamma _{m}}\right\vert _{\vec{\gamma}=\vec{x}}~.
\end{equation}%
The behavior of the system in the (infinitesimal, immediate) vicinity of its
equilibrium configuration is then characterized by exponentials $\exp \left(
\varphi _{m}t\right) $ where the $N$ numbers $\varphi _{m}$ are the $N$
eigenvalues of the $N\times N$ time-independent matrix $\mathbf{F}\left( 
\vec{x}\right) .$ And comparing this behavior with that of the general
solution of the \textit{solvable} system (\ref{EqMotSolvSyst}) one gets
information on the values of these eigenvalues $\varphi _{m}$.

For instance in this manner, by taking as starting point the \textit{solvable%
} dynamical system 
\end{subequations}
\begin{equation}
\dot{\gamma}_{m}\left( t\right) =\mathbf{i~}\left\{ \gamma _{m}\left(
t\right) -\sum_{\ell =1,~\ell \neq m}^{N}\left[ \gamma _{m}\left( t\right)
-\gamma _{\ell }\left( t\right) \right] ^{-1}\right\} ~,  \label{DynSyst}
\end{equation}%
one gets \textit{Diophantine} properties of the matrices (\ref{MHermite})
manufactured with the zeros $x_{n}$ of Hermite polynomials, 
\begin{subequations}
\label{HermiteZeros}
\begin{equation}
H_{N}\left( x_{n}\right) =0~,
\end{equation}%
since these zeros indeed correspond to the equilibrium configuration of the
dynamical system (\ref{DynSyst}) because they satisfy---see for instance 
\cite{St1885} \cite{Sz1975} \cite{C2001}---the $N$ algebraic equations%
\begin{equation}
x_{n}=\sum_{\ell =1,~\ell \neq n}^{N}\left( x_{n}-x_{\ell }\right) ^{-1}~.
\end{equation}%
These \textit{Diophantine} relations consist in the discovery (see for
instance \cite{ABCOP1979} \cite{C2001}) that the $N\times N$ matrix $\mathbf{%
M}\left( \tilde{x}\right) ,$ see (\ref{MHermite}), features the $N$
eigenvalues $0,1,...,N-1$.

The next step is to introduce the monic polynomial with \textit{coefficients}
$\gamma _{m}\left( t\right) $ and \textit{zeros} $\xi _{n}\left( t\right) ,$ 
\end{subequations}
\begin{subequations}
\begin{equation}
p_{N}\left( z;\vec{\gamma}\left( t\right) ;\tilde{\xi}\left( t\right)
\right) =z^{N}+\sum_{m=1}^{N}\left[ \gamma _{m}\left( t\right) ~z^{N-m}%
\right] ~,
\end{equation}%
\begin{equation}
p_{N}\left( z;\vec{\gamma}\left( t\right) ;\tilde{\xi}\left( t\right)
\right) =\prod\limits_{n=1}^{N}\left[ z-\xi _{n}\left( t\right) \right] ~,
\end{equation}%
and the associated dynamical system provided by the identity (\ref{xydot})
together with (\ref{DynSyst}): 
\end{subequations}
\begin{subequations}
\label{DynSystksi}
\begin{eqnarray}
&&\dot{\xi}=-\mathbf{i~}\left[ \sum_{\ell =1,~\ell \neq n}^{N}\left( \xi
_{n}-\xi _{\ell }\right) \right] ^{-1}\cdot   \notag \\
&&\cdot \sum_{m=1}^{N}\left[ \xi _{n}^{N-m}~\left\{ \gamma _{m}-\sum_{\ell
=1,~\ell \neq m}^{N}\left[ \gamma _{m}-\gamma _{\ell }\right] ^{-1}\right\} %
\right] ~,
\end{eqnarray}%
where of course, in the right hand side, the \textit{coefficients} $\gamma
_{n}\equiv \gamma _{n}\left( t\right) $ should be replaced by their
expressions in terms of the \textit{zeros} $\xi _{n}\equiv \xi _{n}\left(
t\right) ,$%
\begin{equation}
\gamma _{m}=\left( -1\right) ^{m}~\sigma \left( \tilde{\xi}\right) ~.
\end{equation}

We look then at the time evolution of the solutions $\xi _{n}\left( t\right) 
$ of this dynamical system in the immediate vicinity of its equilibrium
configuration $\gamma _{m}\left( t\right) =\gamma _{m}\left( 0\right) =x_{m}$%
, $\xi _{n}\left( t\right) =\xi _{n}\left( 0\right) =x_{n}^{\left( \mu
_{1}\right) }$. It is then easily seen, by setting, together with (\ref%
{gammaw}), 
\end{subequations}
\begin{equation}
\xi _{n}\left( t\right) =x_{n}^{\left( \mu _{1}\right) }+\varepsilon
~v_{n}\left( t\right) ~,~~~\tilde{\xi}\left( t\right) =\tilde{x}^{\left( \mu
_{1}\right) }+\varepsilon ~\tilde{v}\left( t\right) ~,
\end{equation}%
and by proceeding as above,\thinspace that one arrives at the linear
evolution system 
\begin{equation}
\overset{\cdot }{\tilde{v}}=\mathbf{M}^{\left( \mu _{1}\right) }\left( 
\tilde{x},\tilde{x}^{\left( \mu _{1}\right) }\right) \mathbf{~}\tilde{v}~,
\end{equation}%
where the $N\times N$ matrices $\mathbf{M}^{\left( \mu _{1}\right) }\left( 
\tilde{x},\tilde{x}^{\left( \mu _{1}\right) }\right) $ are defined by (\ref%
{M1}) (with (\ref{MHermite}) and of course the matrices $\mathbf{R}\left( 
\tilde{x}\right) $ and $\mathbf{\left[ \mathbf{R}\left( \tilde{x}\right) %
\right] ^{-1}}$ defined as above, see (\ref{Id3})), the numbers $x_{n}$ are
the $N$ zeros of the Hermite polynomial $H_{N}\left( z\right) $ of order $N$
and the numbers $x_{n}^{\left( \mu _{1}\right) }$ are the $N$ \textit{zeros}
of the (monic) polynomials the \textit{coefficients} of which are the 
\textit{zeros} of the Hermite polynomial $H_{N}\left( z\right) $ ordered
according to the their permutation $\mu _{1}$.

It is plain from the treatment sketched in this Appendix B that the
statements reported in the last part of Section 4 are validated.

\end{document}